# Viscoelastic dynamics of nanoparticles optically trapped in moving fringe pattern in air-filled hollow-core fiber

SOUMYA CHAKRABORTY,[1,2] GORDON K. L. WONG,[2] PHILIP ST.J. RUSSELL,[1,2] AND NICOLAS Y. JOLY[2,1,†]

[1]*Max Planck Institute for the Science of Light, 91058 Erlangen, Germany*
[2]*Department of Physics, Friedrich-Alexander-Universität, 91058 Erlangen, Germany*
*Corresponding author(s):* [†]nicolas.joly@mpl.mpg.de

We report optical trapping and transport of nanoparticles in a moving interference pattern in hollow-core photonic crystal fiber at atmospheric pressure, when competition between trapping and drag forces causes the particle velocity to oscillate as it is momentarily captured and accelerated by each passing fringe, followed by release and deceleration by viscous forces. As a result the average particle velocity is lower than the fringe velocity. We refer to this phenomenon as "drag-trapping". An analytical model of the resulting motion shows excellent agreement with experiment. Additional control is possible by introducing an imbalance in the backward and forward powers. The high precision of this new technique makes it of interest for example in characterizing nanoparticles, exploring viscous drag forces in different gases and liquids, and temperature sensing.

## 1. INTRODUCTION

Pioneered by Arthur Ashkin [1], laser tweezers has become an essential tool for trapping and manipulating micron-scale objects, and has contributed substantially to scientific progress in many fields, including medicine, biophysics, metrology, and fundamental physics [2,3]. The technique is however unsuitable for transporting small particles over long distances, because a tight spot can be maintained only over a short distance at the focus of a lens. Although diffraction-free Bessel beams, in which the central lobe does not broaden with distance, provide a partial solution, they have high leakage loss [4], as do hollow capillaries [5]. Hollow-core photonic crystal fiber (HC-PCF), on the other hand, by offering ultralow loss diffraction-free propagation in a single-lobed fundamental mode, makes it possible to trap a particle radially by gradient forces, and propel it by radiation pressure over long distances, even along a curved path [6]. As well as offering tight control of parameters, the environment surrounding the levitated particle can be altered by evacuating the core or filling it with gases or liquids. By adding a second backward-propagating mode, the trapped particle may be brought to a halt at any position along the fiber, which can be km long. Such in-fiber levitated particles have been used as point sensors of temperature and electric field, offering a spatial resolution of order the particle diameter $d$ (typically a few µm) over ultralong distances [7]. When $d = \sim\lambda/(2n)$ or less, where $\lambda$ is the vacuum wavelength and $n$ the modal index, it is possible to strongly trap nanoparticles in the interference pattern that forms between counter-propagating laser beams. By varying the relative phase between the counterpropagating modes or introducing a frequency difference, the fringes can be made to move along the fiber, forming a particle conveyor belt, as has been demonstrated in HC-PCF under high vacuum [8,9].

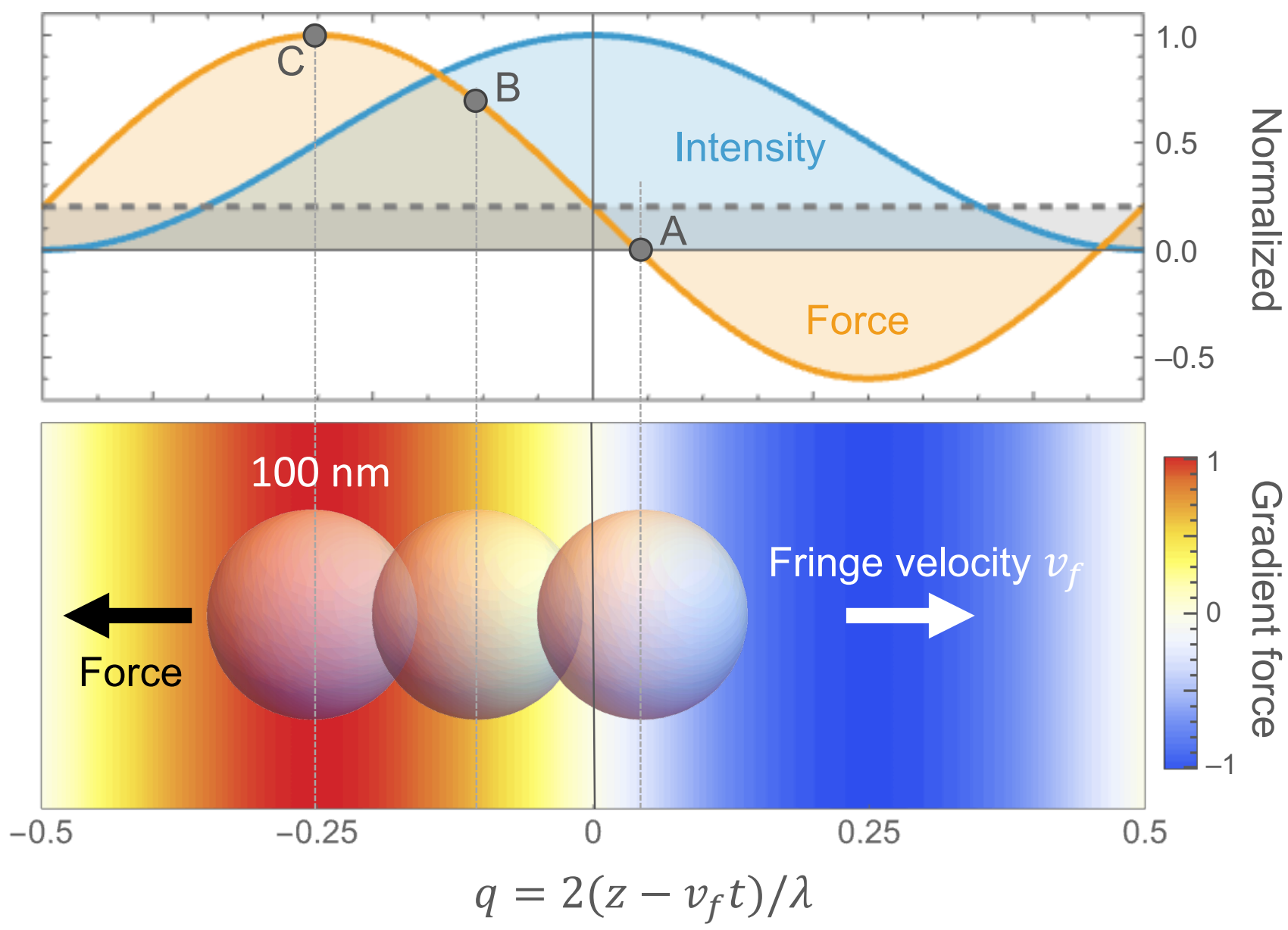

Fig. 1: Effect of viscous drag and radiation pressure (caused by imbalance in backward and forward powers) on the position of a particle trapped within a fringe, as the fringe moves to the right. Upper: fringe intensity and trapping force as a function of position, normalized to fringe width $\lambda/2$ in a frame moving at the fringe velocity $v_f$. The unbalanced radiation force pushes the particle away from the center of the fringe. Lower: particle trapped in fringe. The ratio of fringe width to particle diameter (532/100) is drawn to scale for a 100 nm particle. For stationary fringes the particle is trapped at point A. As the fringe velocity increases the particle is dragged backwards by viscous forces (B). At a certain critical condition (C) the trapping force is exceeded and the particle escapes and is captured by the next fringe (see section 3.3 for details).

Here we report a novel approach to manipulating levitated nanoparticles in HC-PCF, involving the interplay between viscous drag, optical trapping forces and radiation pressure in a moving interference pattern at atmospheric pressure. Under these conditions viscous forces drag the trapped nanoparticle away from the fringe center as the fringe velocity increases or the radiation pressure is unbalanced (see Fig. 1). Above a certain critical velocity, which depends on the pressure, the particle escapes, decelerates, and is captured by the next fringe, where it is again accelerated before escaping into the next fringe, and so on. The result is an average particle velocity that is less than the fringe velocity. By adjusting the trapping power, gas pressure, and relative phase between the trapping modes, the position and velocity of a trapped nanoparticle can be precisely controlled. In this paper we explore this effect experimentally in detail, and develop a nonlinear dynamical theory to explain the results.

## 2. EXPERIMENTAL SET-UP

The HC-PCF structure (Fig. 2c) consists of a single ring of thin-walled capillaries mounted inside a large thick-walled capillary and surrounding a central hollow core. It was drawn from a spinning preform [10], yielding a chiral structure with a twist rate $\alpha = 0.505$ rad/mm. This resulted in circular birefringence $B_C = \sim 10^{-8}$ and the preservation of circular polarization states, causing the electric field of a linearly polarized mode to smoothly rotate at a rate of ~0.17° per m as it travels along the fiber [11]. The fiber loss (1.2 dB/m) was negligible over the ~10 cm fiber length used in the experiments. The laser light (1064 nm continuous wave, linearly polarized) had a linewidth of ~20 kHz and a coherence length of ~5 km, ensuring that the counter-propagating modes interfere with high visibility once the launched polarization states are adjusted so that the electric fields are parallel at the trapping site. Laser light was coupled into both fiber ends using lenses of focal length 75 mm. The balance between backward and forward powers was adjusted using a combination of half-wave plates and polarizing beam splitters (Fig. 2a).

Silica nanoparticles (Kisker Biotech GmbH) were supplied as a suspension in an isopropanol/water mixture with a mass concentration of 50 mg/ml. The manufacturers state that the nominally 195 nm particles range in diameter from 180 to 220 nm. We assume the silica has a refractive index of 1.45 and a density of 2200 kg/m$^3$. The suspensions were diluted in isopropanol to a concentration of 0.015 particles per μm$^3$, corresponding on average to one particle per mist droplet 5 μm in diameter, produced by an Omron nebulizer with mesh size 7 micron.

Droplets, each containing a single silica nano-particle, were laser tweezered just outside the fiber endface until all the liquid had evaporated, and then the particle was launched into the core. Care was taken to position the particle in an unobstructed section within the fiber, free of perturbations caused by scattering defects or mounting clamps. Particle motion was monitored through the fiber cladding using a lens of 60 mm focal length to collimate the side-scattered trapping light, which was then expanded 2× and split into two parts, one delivered to a fast video camera (MotionBLITZ EoSens mini2) and the other to an InGaAs amplified photodiode (Thorlabs PDA20C2).

The relative phase of the trapping beams, and thus the position and velocity of the fringes, was controlled using a Pockels phase modulator (Thorlabs EO-PM-NR-C2), driven by a combination of signal generator (NI-DAQ USB 6251) and high-voltage amplifier (New Focus 3211). A steady unidirectional fringe velocity $v_f$ was synthesized by driving with a sawtooth wave of peak-to-peak amplitude 2π, yielding $v_f = f\lambda/2$ where $f$ is the sawtooth frequency and $\lambda$ the vacuum wavelength. To ensure that the propulsive radiation pressure on the particle, which is separate from the gradient forces, was perfectly balanced in the experiment, we measured the particle velocities for positive and negative sawtooth ramps and verified that they were equal and opposite.

A typical measurement of particle position versus time is shown in Fig. 2d, where the particle velocity is 2.2 mm/s, compared to a fringe velocity of 4.79 mm/s, and confirming that drag-trapping is taking place.

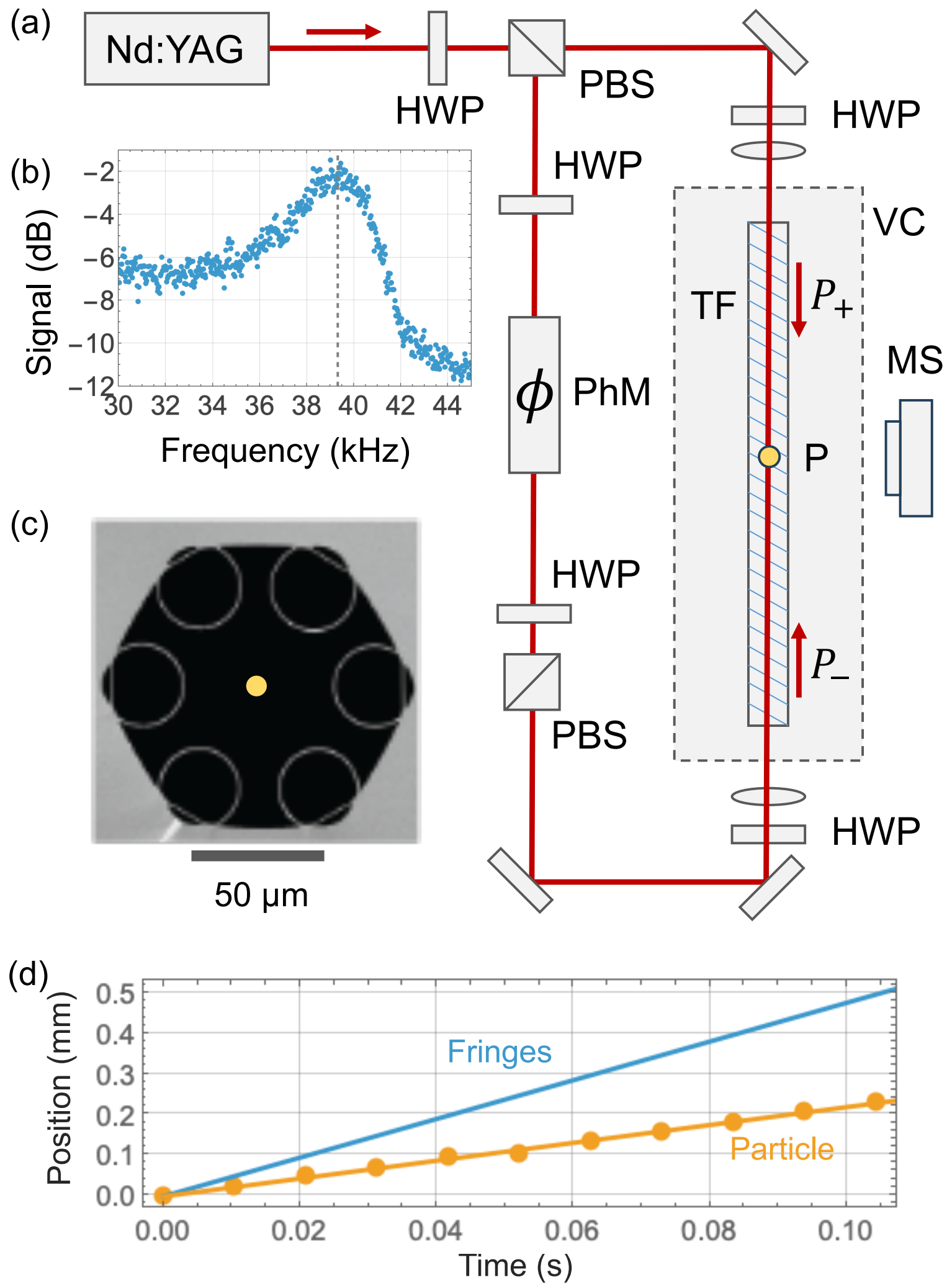

Fig. 2: (a) Experimental setup: HWP: half-wave plate, PBS: polarizing beam splitter, MS: motion sensor, PhM: phase modulator, TF: twisted HC-PCF, P: particle (silica nanosphere), VC: vacuum chamber. The laser emits continuous-wave light at 1064 nm wavelength. A chiral hollow-core fiber is placed within a vacuum chamber to control the pressure. The two HWP/PBS combinations are used to adjust the backward and forward polarization states and power levels. The total power is kept constant at $P_{tot} = P_+ + P_-$, and a radiation imbalance can be introduced by setting $P_+ \neq P_-$. (b) Measured power spectral density of the thermally-driven axial motion at 12 mbar pressure for a 295 nm particle trapped in a stationary fringe at a total trapping power of 1 W. An axial resonance is visible at ~39.3 kHz. (b) Scanning electron micrograph of the microstructure of the chiral single-ring photonic crystal fiber, hollow core diameter ~44 µm. (c) Raw positional data (orange data-points and fitted curve) for a 195 nm particle at a fringe velocity of 4.79 mm/s (blue line) and atmospheric pressure, with total trapping power 1 W. The particle travels at 2.2 mm/s, slower than the fringes.

## 3. THEORY

### 3.1 Trapping forces & frequencies: stationary fringes

The Poynting vector distribution in the fundamental mode of a HC-PCF approximates to $S(r) = S_0 J_0^2(2u_{01}r/d_{co})$ where $d_{co}$ is the core diameter, $r$ the radial position, $u_{01}$ the first zero of the Bessel function $J_0$, and $S_0$ the value at the center of the core, where the particle is trapped; the electric field intensity is then given by $|E(r)|^2 = 2S(r)/(\varepsilon_0 c)$. By integrating $S(r)$ over the core we find $S_0 = 4P/(\pi d_{co}^2 J_1^2(u_{01})) \cong 4.7P/d_{co}^2$ where $P$ is the power in a unidirectional LP01-like core mode. When two counterpropagating modes with unbalanced powers interfere, keeping the total power constant, i.e., $P_{tot} = P_+ + P_-$, the electric field time-averaged energy density can be written in the form:

$$\rho_e(r,z) = \frac{S(r)}{c}\left|\sqrt{2}\sin\left(\frac{\pi}{4}+\frac{\delta}{2}\right)e^{i\beta z} + \sqrt{2}\cos\left(\frac{\pi}{4}+\frac{\delta}{2}\right)e^{-i\beta z}\right|^2 = \frac{2S(r)}{c}(1+\cos\delta\cos 2\beta z) \tag{1}$$

where $\beta = 2\pi/\lambda$ is the propagation constant and $c$ the speed of light in vacuum; the power in the forward direction is then $P_+ = P_{tot}\sin^2(\delta/2 + \pi/4)$: for perfect fringe visibility $\delta = 0$, and for zero visibility $\delta = \pm\pi/2$. In the Rayleigh regime ($d < \sim\lambda/20$) the resulting gradient forces can be written in the form $\mathbf{F}_G = (\pi d^3 \xi/2)\nabla\rho_e(r,z)$ where $\xi = (n_s^2 - 1)/(n_s^2 + 2)$ and $n_s$ is the index of silica [12]. Deviations from $\delta = 0$ will introduce an imbalance in radiation pressure, which will push the particle away from the intensity maximum. To model this, we add a radiation force that is proportional to the intensity difference, yielding a total on-axis force in the $z$-direction:

$$F_z = \frac{S_0 \pi d^2}{2c}\left(\kappa_R \sin\delta - \frac{8\pi\kappa_{Gz}\xi d}{\lambda}\cos\delta\sin 2\beta z\right) \tag{2}$$

where $\kappa_R$ and $\kappa_{Gz}$ are correction factors for the radiation and axial gradient forces (necessary because the Rayleigh approximation overestimates the gradient force, and only a small fraction of the incident momentum is converted into a radiation force). These factors can be estimated experimentally (section 4.4). We see that $F_z = 0$ at the trapping position:

$$z = z_0 = \frac{\lambda}{4\pi}\sin^{-1}\left(\frac{\kappa_R \lambda \tan\delta}{8\pi\kappa_{Gz}\xi d}\right) \tag{3}$$

and that axial trapping is possible only if $|\delta| < \delta_{max} = \tan^{-1}[8\pi\kappa_{Gz}\xi d/(\kappa_R\lambda)]$. When $|\delta| > \delta_{max}$ the radiation pressure overcomes over the gradient trapping force, and the particle will oscillate in velocity as it is pushed through successive fringes (section 3.3).

For pure radial and axial trapping at $(r,z) = (0,0)$ we find:

$$\omega_{0r}^2 = \kappa_r \frac{48 u_{01}^2 \xi S_0}{c\rho_s d_{co}^2}, \quad \omega_{0z}^2 = \kappa_{Gz}\frac{96\pi^2 \xi S_0}{c\rho_s \lambda^2}\cos\delta \tag{4}$$

where $\omega_{0r}/2\pi$ and $\omega_{0z}/2\pi$ are the small-signal resonant frequencies of the trapped particle and $\rho_s$ is the density of silica. For $d = 195$ nm and $d_{co} = 44$ µm we find $\omega_{0z}/\omega_{0r} = \sim 50$, using $\kappa_r = 0.59$ and $\kappa_z = 0.25$ (obtained by comparison with measured resonant frequencies at 0.25 mbar, see section 4.4). The amplitude $z_m$ of the thermally driven axial vibrations at low pressure can be estimated using the equipartition theorem, $k_B T/2 = k_s z_m^2/2$, where $k_s = \partial F_z/\partial z$, $k_B$ is Boltzmann's constant and $T$ the absolute temperature, yielding:

$$|z_m| = \sqrt{\frac{6k_B T}{\omega_{0z}^2 \rho_s \pi d^3}}. \tag{5}$$

For example, $d = 195$ nm, $\omega_0/2\pi = 51.4$ kHz and $T = 293$ K gives $|z_m| = 67$ nm; and $d = 832$ nm, $\omega_0/2\pi = 20.0$ kHz and $T = 293$ K gives $|z_m| = 20$ nm.

### 3.2 Viscous drag force

The viscosity of air as a function of absolute temperature $T$ follows Sutherland's law [13]:

$$\mu_0(T) = \mu_{\text{ref}}\left(\frac{T}{T_{\text{ref}}}\right)^{3/2}\frac{T_{\text{ref}} + S}{T + S} \tag{6}$$

where $T_{\text{ref}} = 293$ K, $S = 111$ K and $\mu_{\text{ref}} = 1.84 \times 10^{-5}$ Pa.s. The drag coefficient $C_d$ for a particle may be written in the form [7,14,15]:

$$C_d = \frac{F_d}{v_p} = K_c(p)3\mu_0\pi d = \frac{3\mu_0\pi d}{1 + K_n(p)(\beta_1 + \beta_2 e^{-\beta_3/K_n(p)})} \quad (7)$$

where $F_d$ is the drag force, $K_c$ is a pressure-dependent correction factor, $d$ is the particle diameter, $v_p$ the particle velocity, $p$ is the pressure in Pa, $\beta_1 = 1.231$, $\beta_2 = 0.469$ and $\beta_3 = 1.178$. The Knudsen number $K_n$ is:

$$K_n(p) = \frac{\Lambda}{d_c} = \frac{k_B T/p}{d_c\sqrt{2}\pi d_m^2} \quad (8)$$

where $\Lambda$ is the molecular mean free path, $d_c$ is a characteristic length, $k_B$ is Boltzmann's constant, $p$ the pressure in Pa, $d_m = 3.6$ Å the diameter and $a_c = \pi d_m^2/4$ the collision cross-section of the molecules, and $\rho_N$ their number density. We found that $d_c = d/2.371$ gave good agreement with the measurements for the particle diameters tested (see below).

### 3.3 Dynamics of motion

The axial motion of a nanoparticle trapped in a moving fringe pattern can be modelled as a driven damped nonlinear harmonic oscillator:

$$m\ddot{q}/C_d + \dot{q} + v_{crn}\sin(2\pi q) = v_{Rn} - v_{fn} \quad (9)$$

where $q = 2(z - v_f t)/\lambda$ is the particle position in a frame moving at the normalized fringe velocity $v_{fn} = 2v_f/\lambda$, $z$ is the axial position, $m$ is the particle mass, $C_d$ N.s/m is the drag coefficient, and $v_{Rn} = S_0\pi d^2\kappa_R/(c\lambda C_d)\sin\delta$ is the normalized velocity associated with a radiation pressure imbalance. In the steady-state at constant particle velocity, $q = (2\pi)^{-1}\sin^{-1}[(v_{Rn} - v_{fn})/v_{crn}]$, showing that the particle reaches the trap edges $q = \pm 1/4$ when $(v_{Rn} - v_{fn}) = \pm v_{crn}$, where $v_{crn} = m\omega_{0z}^2/(2\pi C_d)$ is the normalized critical fringe velocity at which the particle is dragged to the edge of the trap by viscous forces. For $(v_{Rn} - v_{fn}) > v_{crn}$ there is no stable trapping position, and the particle slips from fringe to fringe. Although Eq.(9) can readily be solved numerically, a highly accurate analytical solution can be found at atmospheric pressure when the damping is very high, so that the term $m\ddot{q}/C_d$ can be neglected, leading to the solution:

$$q(t) = -\pi^{-1}\tan^{-1}\left(\zeta_R + (1 - \zeta_R^2)^{1/2}\ \tan\left[\pi t(v_{fn} - v_{Rn})(1 - \zeta_R^2)^{1/2}\right]\right) \quad (10)$$

where $\zeta_R = v_{crn}/(v_{fn} - v_{Rn})$. We see that oscillating solutions exist only when $|\zeta_R| < 1$, with frequency $\left((v_{fn} - v_{Rn})^2 - v_{crn}^2\right)^{1/2}$. For $|\zeta_R| \geq 1$ the particle is trapped inside a fringe and so the frequency is zero. Note also that at low pressure the $\ddot{q}$ term can no longer be neglected so Eq.(9) must be solved numerically.

Theoretical position-versus-time plots of $q(t)$ (Fig. 3), in a frame moving with the fringes so that the particle velocity is negative, show how in each case the particle is captured by a fringe, held for a moment while the drag force pulls it to the edge of the trap, and then released and captured by the next fringe.

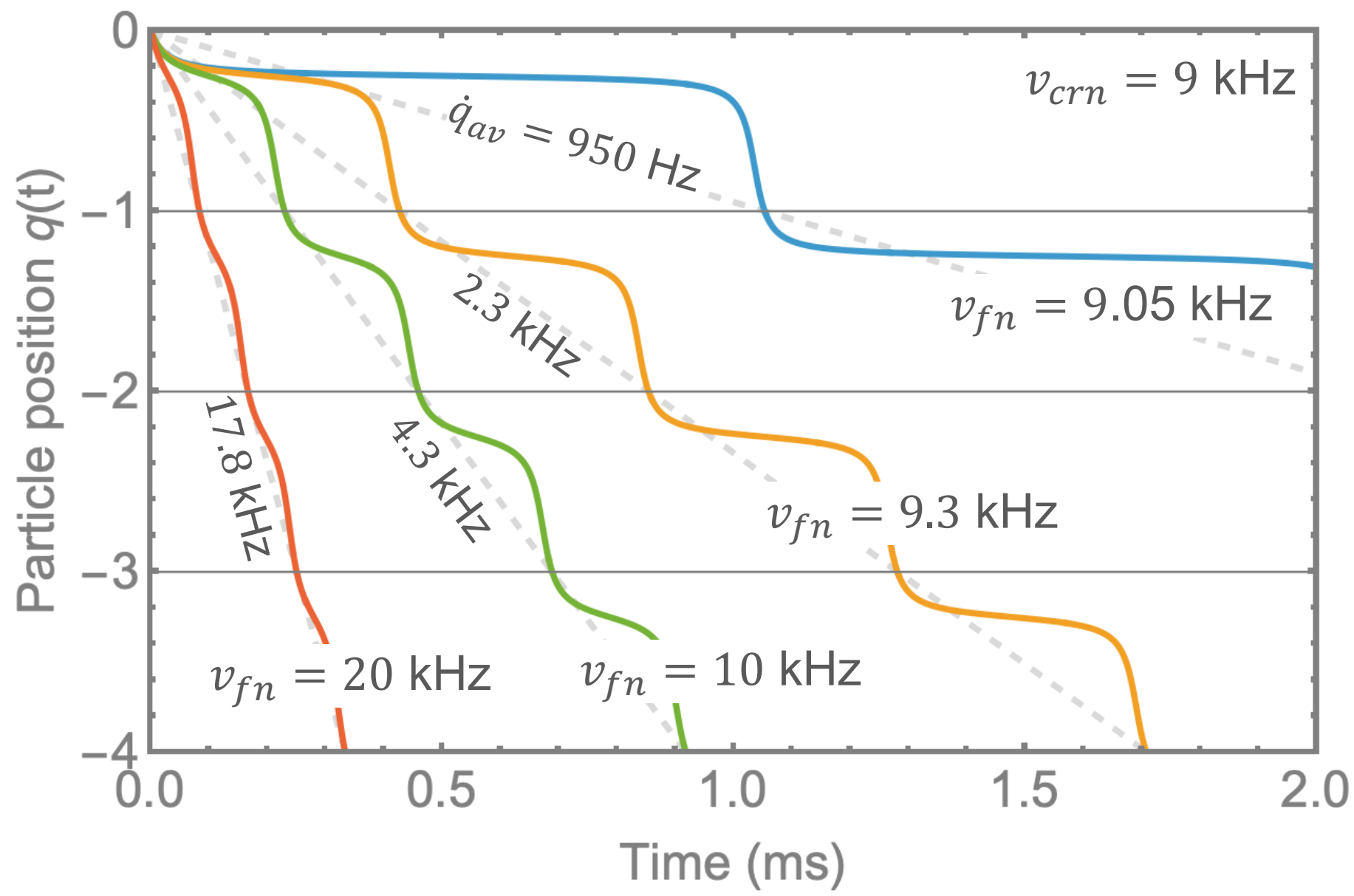

Fig. 3: Normalized particle position versus time for $v_{crn} = 9$ kHz, corresponding to $d = 195$ nm and $P_{tot} = 1$ W (see Fig. 6) at atmospheric pressure, for four values of $v_{fn}$ and $v_{Rn} = 0$. The curves correspond to $\zeta_R = v_{crn}/v_{fn} =$ 0.994, 0.968, 0.93 and 0.6. As the fringe velocity increases, the particle slips more and more rapidly from fringe to fringe, while the motion becomes more and more sinusoidal. The dashed lines mark the average particle slip-back velocity, $\dot{q}_{av} = (v_{fn}^2 - v_{crn}^2)^{1/2}$, labelled with slanted text.

### 3.4 Side-scattered signals

The simplest way to follow the particle motion is by detecting the side-scattered light signal. At low pressure the resonant motion of a trapped particle can be directly measured, and at atmospheric pressure the periodic motion of the particle can be monitored as it moves through the fringe pattern. In general, the side-scattered signal will take the form $I_{ss}(t) = I_0 \cos^2[\pi q(t)]$ where $q(t)$ describes the motion of the particle relative to the intensity pattern. As $I_{ss}(t)$ is periodic in time in both cases, it can be represented as a Fourier series:

$$I_{ss}(t) = \sum_{p=-\infty}^{\infty} a_p e^{i2\pi pt/T}, \qquad a_p = \frac{1}{T}\int_0^T I_0 \cos^2[\pi q(t)]\, e^{-i2\pi pt/T}\, dt \tag{11}$$

where $p$ is the harmonic order and $T$ is the period, which equals $2\pi/\omega_{0z}$ for stationary fringes at low pressure, and $1/\left[(v_{fn} - v_{Rn})^2 - (v_{crn}\cos\delta)^2\right]^{1/2}$ for moving fringes at atmospheric pressure. The $p$-th mean-square harmonic intensity at a square-law detector is then $2a_{+p}a_{-p}$, where $a_{-p} = a_{+p}^*$.

## 4. EXPERIMENTAL MEASUREMENTS

### 4.1 Thermally driven vibrations at low pressure

The intensity of the thermally-driven signal will vary as $I_{ss}(t) = I_0 \cos^2[(\pi(q_0 + q_m \sin\omega_{0z}t)]$ where $q_m$ is the amplitude of the vibrations. In the absence of any power imbalance, $q_0 = 0$ and the side-scattered signal will be weak and at frequency $2\omega_{0z}$. As $|q_0|$ increases from zero, however, a stronger and stronger signal will appear at frequency $\omega_{0z}$, until $\delta = \sim\pi/4$. It is thus beneficial to introduce a degree of power imbalance to enhance the signal at $\omega_{0z}$. Since the particle is not a point scatterer, however, the finer features in the signal will tend to be smoothed out in the experiment. Experimental measurements of $\omega_{0z}$ are given in section 4.4 and in Fig. 2(b).

### 4.2 Particle slipping through fringes

The signal detected in side-scattering, $I_{ss}(t)$, will oscillate in strength as the particle moves through the fringes. It can be modelled using the function $q(t)$:

$$I_{ss}(t) = I_0 \cos^2[\pi q(t)] \quad (12)$$

where $q(t)$ is given in Eq.(10), $I_0$ scales with the trapping power, and we make the assumption that the signal is roughly proportional to the fringe intensity at position $q(t)$. Although there is considerable spectral broadening due to Brownian motion, a clear peak is seen at ~18 kHz for $v_{fn} = 20$ kHz (orange data). For $v_{fn} = 10$ kHz, however, the trapping is stronger, the repetition rate is ~4.3 kHz, and the highly non-sinusoidal particle motion produces many harmonics, which are smoothed out by Brownian noise (blue data). Nevertheless the agreement between theory and experiment is good.

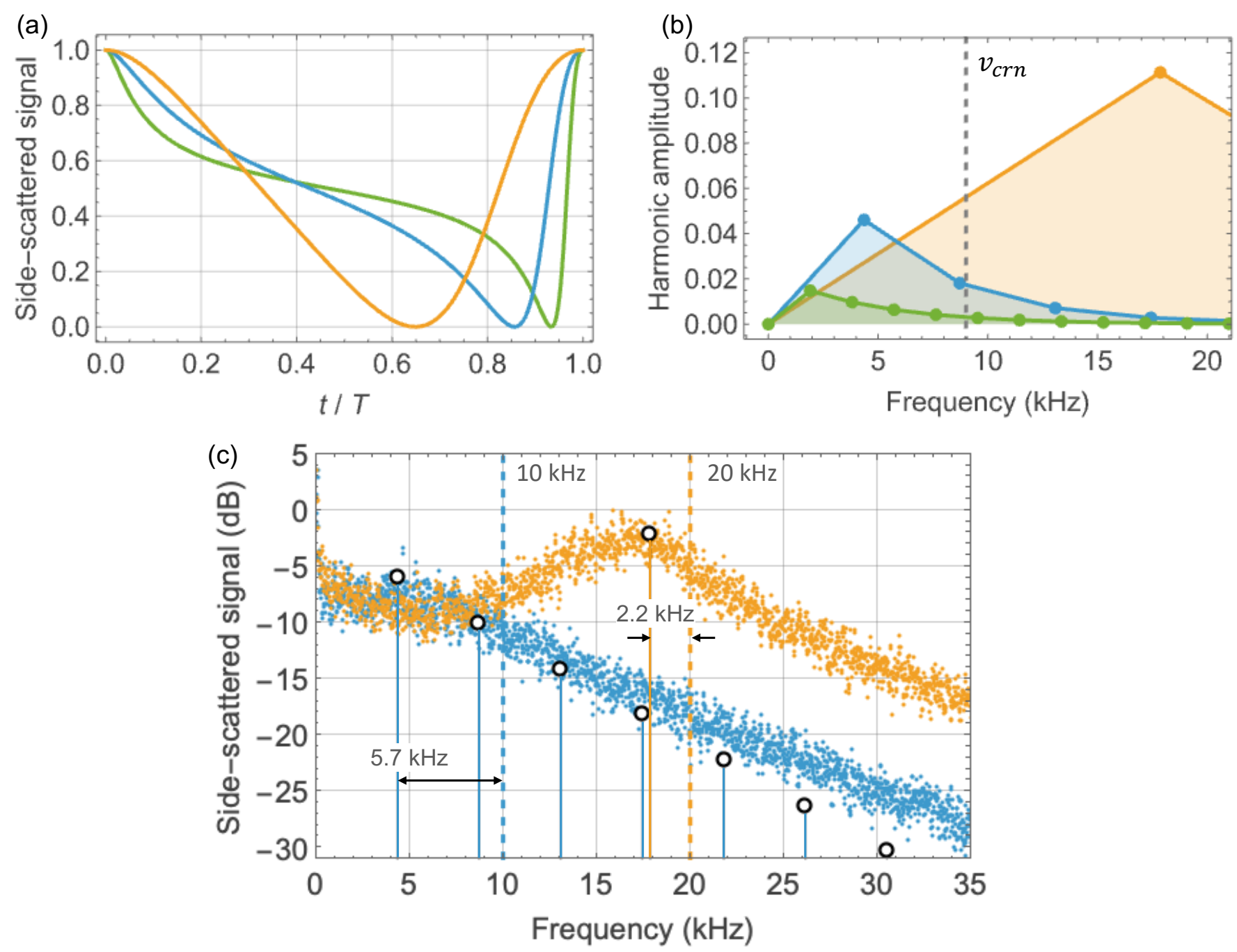

Fig. 4: (a) Calculated side-scattered signal over one period $T = (v_{fn} - v_{Rn})^{-1}(1 - \zeta_R^2)^{-1/2}$ for $\delta = 0$, $v_{crn} = 9$ kHz, $v_{Rn} = 0$, and $v_{fn} = 9.2$ kHz (green), 10 kHz (blue) and 20 kHz (orange). (b) Harmonic amplitudes for these three cases (data-points, calculated by a Fourier series, Eq.(11)), with the dc term set to zero. As $v_{fn}$ increases, the repetition rate rises and the fundamental harmonic becomes stronger, while the side-scattered signal becomes more sinusoidal. (c) Spectra of the measured side-scattered signals for $v_{crn} = 9$ kHz, compared with theory (black circles). For $v_{fn} = 20$ kHz (orange) a clear first harmonic is seen at ~17.8 kHz (which corresponds with the slip-back frequency), as expected of a more sinusoidal signal. For $v_{fn} = 10$ kHz (blue), however, the signal is highly non-sinusoidal, producing many more harmonics, with a slip-back frequency of ~4.3 kHz. In both cases there is good agreement between experiment and theory.

### 4.3 Correction factors for gradient and radiation forces

Correction factors for the Rayleigh gradient force are defined in Eq.(4). These are given by $\kappa_{Gr} = (\omega_{0r}/\omega_{0r}^R)^2$, $\kappa_{Gz} = (\omega_{0z}/\omega_{0z}^R)^2$ and can be calculated by comparing the with the Rayleigh frequency with the measured resonant frequency at low pressure. We directly measured thermally-driven axial and radial side-scattered spectra for a 195 nm particle at 1 W total trapping power, $v_f = 0$, and pressure 0.25 mbar. Resonances are seen at $\omega_{0r}/2\pi = 1.02$ kHz and $\omega_{0z}/2\pi = 51.5$ kHz. When interference was suppressed by making the trapping electric fields orthogonal, the axial resonance disappeared and the radial frequency dropped by a factor of $\sim\sqrt{2}$ to 900 Hz, as expected of a twice smaller on-axis electric field intensity.

Measured resonant frequencies and correction factors for several different particle sizes are plotted in Fig. 5(b&c). Also included is the correction factor for the radiation pressure $\kappa_R$. In Fig. 2 a resonance was detected at 39.3 kHz for a 295 nm particle at a total balanced trapping power of 1 W. Evaluating Eq.(4) we find $\omega_{0z}/2\pi = 39.7$ kHz using $\kappa_{Gz} = 0.15$ from Fig. 5(c), in good agreement with the measured frequency.

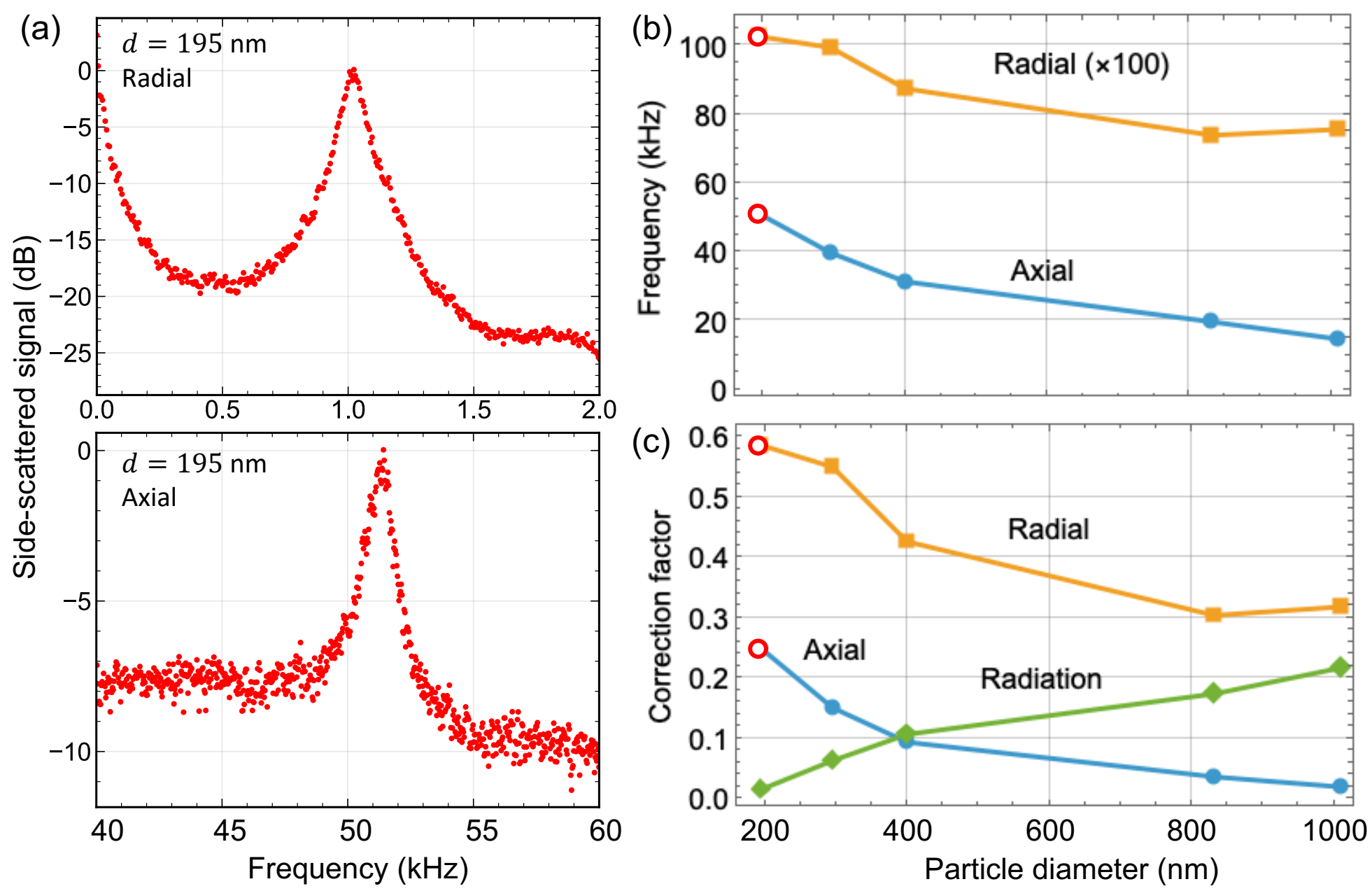

Fig. 5: (a) Measured axial and radial resonances in stationary fringe-trap at 1 W total trapping power for 195 nm diameter particle at pressure < 1 mbar. The signal can be optimised by introducing a degree of imbalance in the trapping powers. (b) Measured radial and axial resonant frequencies for a total 1 W trapping power and several different particle sizes at pressure < 1 mbar. (c) Gradient force correction factors $\kappa_G = (\omega_0/\omega_0^R)^2$ (right) in axial and radial directions for a range of different particle diameters at 1 W total trapping power. Also included is the radiation force correction factor $\kappa_R$ for a unidirectional beam, yielding $F_{Rz} = \kappa_R S_0 \pi d^2/4c$ (see section 3.1).

Using the video camera, we measured the average velocity of a 195 nm particle at a range of different trapping powers; the data is presented in Fig. 6. When the particle is trapped in a fringe, $v_p = v_f$, which corresponds to the slanting straight dashed line. For $v_p < v_f$, the particle escapes the trap and oscillates in velocity as successive fringes pass through it. Good agreement with the analytical theory is found, except at lower power when the trap is weaker and more strongly perturbed by Brownian noise, especially in the vicinity of the critical fringe velocity.

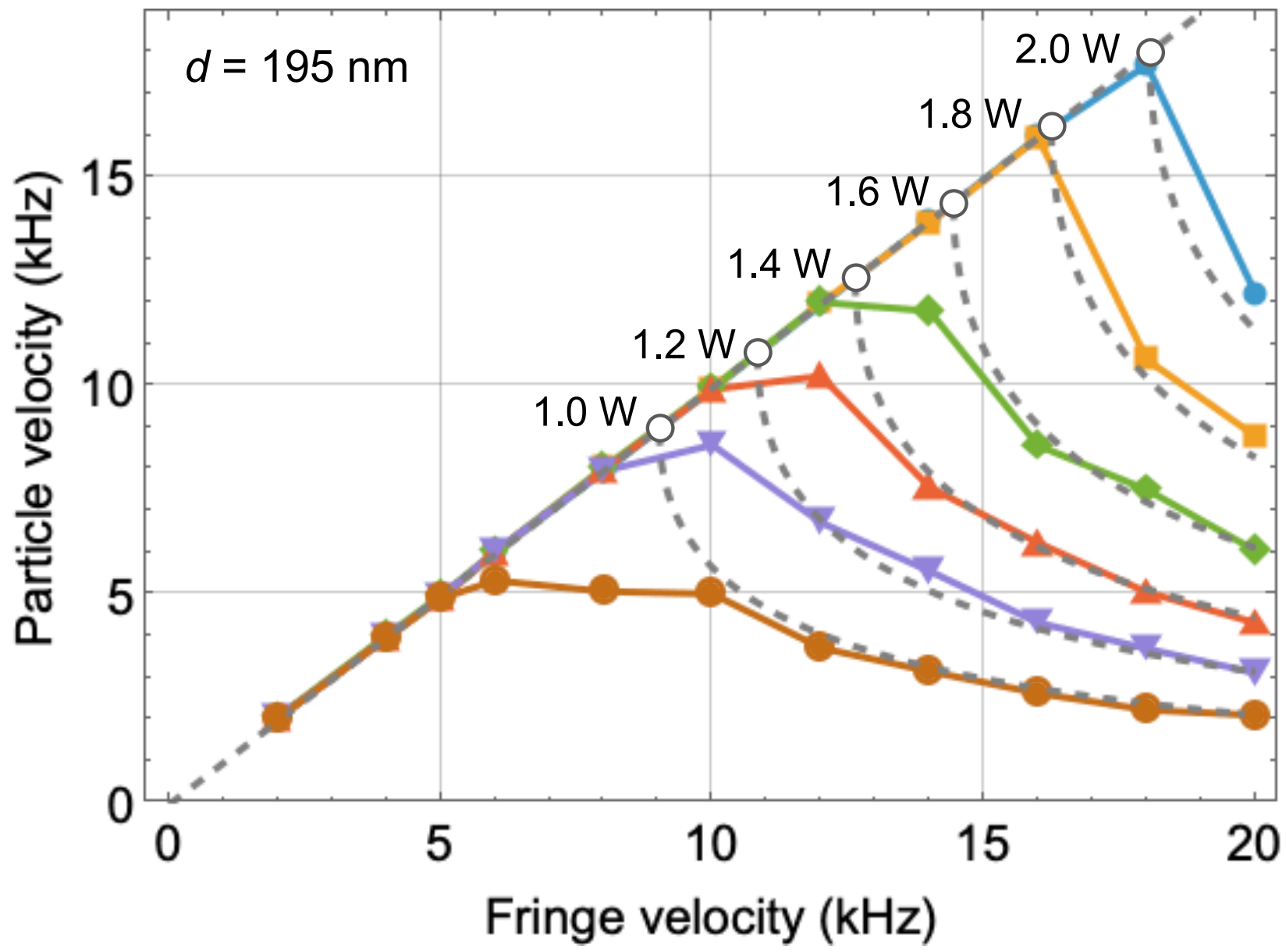

Fig. 6: Comparison of theory and experiment for a particle with $d = 195$ nm and 6 different total power levels ($P_+ = P_-$). The dashed gray curves show the analytical theory, and the white circles mark the theoretical points at which the fringe and particle velocities are both equal to $v_{crn}$; for higher fringe velocities the particle slips out of the trap. The datapoints are experimental measurements, joined by straight lines for clarity. The agreement is better at higher power, when the trap is stronger and more resistant to Brownian motion.

### 4.4 Resonant frequencies in the presence of power imbalance

We confirmed the $\sqrt{\cos\delta}$ dependence of the resonant frequencies in a series of measurements, shown in Fig. 7. The frequency drops with increasing power imbalance, as expected. The slight disparity between theory and experiment we attribute to having used a fresh particle—the particles are only nominally 195 nm in diameter.

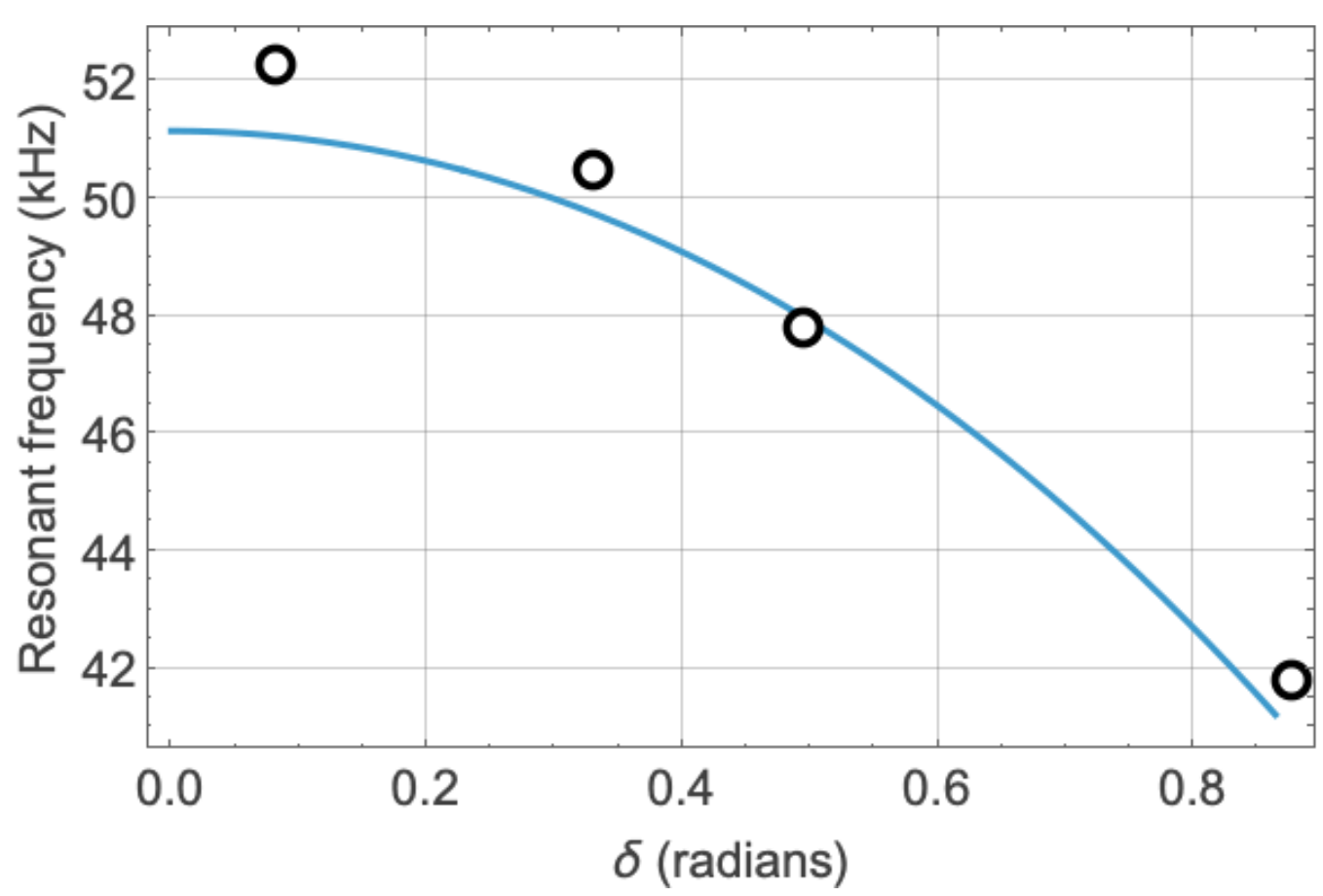

Fig. 7: Measured resonant frequencies of a nominally 195 nm particle at 1 W total trapping power as the power imbalance increases (open circles). The full blue line is the theory from Eq.(4).

## 5. DISCUSSION & CONCLUSIONS

In conclusion, the axial motion of nanoparticles in a HC-PCF can be precisely controlled by trapping them in the fringes of a moving interference pattern. The particles undergo a repetitive cycle of being dragged out of a fringe, decelerated by viscous forces, captured and momentarily accelerated by the next fringe, and so on. As a result the average particle velocity is less than the fringe velocity. The particle velocity can be additionally adjusted by deliberately introducing a power imbalance in the trapping beams. The technique also offers a convenient means of accurately calibrating the optical and viscous forces acting on nanoparticles levitated in different gases and fluids at different pressures and temperatures—difficult to achieve using other approaches. It also offers an alternative means of measuring the diameter and refractive index of nanoparticles, after careful calibration of the optical forces and resonant frequencies for particles of known size and index [16]. Since the air viscosity depends on temperature, drag-trapping has the potential to be used as a thermometer. We plan to study these and other ideas in future work.

## 6. FUNDING, ACKNOWLEDGEMENTS AND AUTHOR CONTRIBUTIONS

The work was funded by the Deutsche Forschungsgemeinschaft and the Max-Planck-Gesellschaft. SC and GKLW carried out the experiments, PR developed the theory, SC and PR produced the fits to the experimental data, and NJ supervised the project.

## 7. CONFLICTS OF INTEREST

The authors declare no conflicts of interest.

## 8. DATA AVAILABILITY STATEMENT

Data underlying the results presented in this paper are not publicly available at this time but may be obtained from the authors upon reasonable request.